# Ultrathin natural biotite crystals as a dielectric layer for van der Waals heterostructure applications

Raphaela de Oliveira[1], Ana Beatriz Yoshida[2,3], Cesar Rabahi[4], Raul O. Freitas[1], Christiano J. S. de Matos[5,6], Yara Galvão Gobato[4], Ingrid D. Barcelos[1], and Alisson R. Cadore[2*]

[1]*Brazilian Synchrotron Light Laboratory (LNLS), Brazilian Center for Research in Energy and Materials (CNPEM), 13083-970, Campinas, SP, Brazil.*

[2]*Brazilian Nanotechnology National Laboratory (LNNano), Brazilian Center for Research in Energy and Materials (CNPEM), 13083-100, Campinas, SP, Brazil*

[3] *"Gleb Wataghin" Institute of Physics, State University of Campinas, 13083-970, Campinas, SP, Brazil*

[4] *Department of Physics, Federal University of São Carlos, 13565-905, São Carlos, SP, Brazil*

[5]*MackGraphe, Mackenzie Presbyterian Institute, 01302907, Sao Paulo, SP, Brazil*

[6]*Mackenzie Engineering School, Mackenzie Presbyterian University, 01302907, Sao Paulo, SP, Brazil*

*Corresponding author: alisson.cadore@lnnano.cnpem.br

Biotite, an iron-rich mineral belonging to the trioctahedral mica group, is a naturally abundant layered material (LM) exhibiting attractive electronic properties for application in nanodevices. Biotite stands out as a non-degradable LM under ambient conditions, featuring high-quality basal cleavage – a significant advantage for van der Waals heterostructure (vdWH) applications. In this work, we present the micro-mechanical exfoliation of biotite down to monolayers (1Ls), yielding ultrathin flakes with large areas and atomically flat surfaces. To identify and characterize the mineral, we conducted a multi-elemental analysis of biotite using energy-dispersive spectroscopy mapping. Additionally, synchrotron infrared nano-spectroscopy was employed to probe its vibrational signature in few-layer form, with sensitivity to the layer number. We have also observed good morphological and structural stability in time (up to 12 months) and no important changes in their physical properties after thermal annealing processes in ultrathin biotite flakes. Conductive atomic force microscopy evaluated its electrical capacity, revealing an electrical breakdown strength of approximately 1 V/nm. Finally, we explore the use of biotite as a substrate and encapsulating LM in vdWH applications. We have performed optical and magneto-optical measurements at low temperatures. We find that ultrathin biotite flakes work as a good substrate for 1L-MoSe$_2$, comparable to hexagonal boron nitride flakes, but it induces a small change of the 1L-MoSe$_2$ g-factor values, most likely due to natural impurities on its crystal structure. Furthermore, our results show that biotite flakes are useful systems to protect sensitive LMs such as black phosphorus from degradation for up to 60 days in ambient air. Our study introduces biotite as a promising, cost-effective LM for the advancement of future ultrathin nanotechnologies.

Keywords: phyllosilicates, naturally abundant, biotite, dielectric breakdown, nanodevices.

## 1 – Introduction

Van der Waals heterostructures (vdWHs) based on stacking different layered materials (LMs) one on top of another [1] have garnered significant interest in recent years due to their potential for designing distinct functionalities in two-dimensional (2D) systems, paving the way for novel studies and applications [2–9]. In pursuit of applications for vdWHs, a wide range of LMs have been investigated, spanning those with high charge carrier mobility [10–12] to the most insulating varieties [13–17], as well as those exhibiting optical [18–21], topological [22–24] or intrinsic/extrinsic magnetic properties in more recent studies [25–28]. Among the insulating LMs, hexagonal boron nitride (hBN) a synthetic layered insulator, has emerged as the most utilized material in vdWH studies [17,29]. However, for the fabrication of nanoelectronic devices based on LMs to realize their full performance potential, it is imperative to expand the selection of suitable insulators to ensure production scalability [17,30]. Consequently, other insulating LMs have been proposed as promising candidates for substrates and capping layers in vdWH applications, such as micas [31–39], clays [16,40–46] and others [13,29,31,47–49].

In this context, most of these alternative insulating LMs explored so far are naturally occurring and belong to the sheet silicate (phyllosilicate) minerals group [13,15]. Increasing research into naturally occurring LMs is highly beneficial because nanomaterials that are abundant on Earth and easy to extract can be incorporated into nanotechnological applications with lower associated costs compared to synthetic alternatives, in addition to being environmentally friendly [15,17,50,51]. This could help ensure the scalability of future nanoelectronics based on LMs. Biotite thus emerges as a particularly interesting phyllosilicate in this search for earth-abundant sources of insulating layered materials.

Biotite (or black mica), with the chemical formula $(Mg,Fe)_3AlSi_3O_{10}(OH)_2$, belongs to the trioctahedral mica group [52]. Recent works have demonstrated that this material exhibits interesting properties for energy harvesting [53,54], while others investigated its mechanical [55], electrical [56,57], optical [58], and magnetic [26] properties in its bulk form. Nevertheless, no systematical study has been conducted to investigate the properties of ultrathin biotite or to demonstrate its used in vdWH applications, despite reports of its mechanical exfoliation [15,26,59,60]. In this work, we present biotite as a non-degradable LM that can be micromechanically cleaved down to a monolayer (1L) with an atomically flat surface, preserving its vibrational characteristics in its few-layer (FL) form. Presenting aging stability under ambient conditions for 12 months, biotite can also withstand annealing in air up to 300°C without undergoing any morphological changes. Its electrical insulating capacity, probed by conductive atomic force microscopy (c-AFM), shows an electrical breakdown strength of ~1 V/nm. Moreover, we demonstrate that biotite can be embedded in different vdWHs as substrate or encapsulating layers. We show that FL-biotite is a suitable substrate for improving the emission quality of neutral excitons (X) from a 1L-MoSe$_2$ in low-temperature photoluminescence (PL) experiments. Furthermore, we also observe that the biotite substrate reduces the exciton g-factor value compared to SiO$_2$ substrates. Additionally, when used as an encapsulating material, biotite can protect sensitive LMs from degradation, such as black phosphorous (BP), for up to 60 days in an open environment. Therefore, we anticipated using biotite crystals in flexible-electronic nanodevices [37], suggesting that embedding naturally occurring LMs in vdWHs can serve as a fundamental building block for potential scalable nanotechnologies [30,61].

## 2 – Methods

### 2.1 – Few-layer biotite exfoliation and assembly of van der Waals heterostructures

The bulk biotite crystal was obtained from Itabira city, Minas Gerais/Brazil. The FL-biotite flakes used in this work were obtained from standard micro-mechanical exfoliation [10] of the biotite-bulk crystal atop (300 nm) SiO$_2$/Si for the atomic force microscopy (AFM) analysis and (100 nm) Au/Si substrates for the c-AFM and synchrotron infrared nanospectroscopy (SINS) analysis [62].

1L-MoSe$_2$/biotite heterostructures were fabricated as follows: i) FL-biotite flakes atop SiO$_2$/Si substrates were selected under the optical microscope and examined by AFM to probe thickness and surface quality. ii) 1L-MoSe$_2$ flakes were obtained from the exfoliation of the MoSe$_2$-bulk crystal (2D Semiconductors) onto a polydimethylsiloxane (PDMS) stamp by scotch tape technique [63]. iii) the 1L-MoSe$_2$/PDMS stamps were then aligned and stamped on the chosen FL-biotite slabs using a commercial transfer system (HQ Graphene).

For assembling the biotite/BP heterostructures we performed the processes inside a glove box under an $N_2$ environment (<2 ppm $O_2$ and $H_2O$): iv) BP-bulk crystal (HQ Graphene) was mechanically exfoliated by scotch tape onto $SiO_2/Si$ substrates. Areas of the substrate with several BP flakes were then selected by optical microscopy. v) FL-biotite flakes were obtained from the exfoliation of the biotite-bulk crystal onto the PDMS stamps, and the selected flakes were examined under the optical microscope. vi) the FL-biotite/PDMS stamps were then aligned and stamped on the chosen BP regions using a homemade transfer system inside the glovebox. Only after the final assembly, the biotite/BP heterostructures were taken out from the glovebox and exposed to ambient conditions.

*2.2 – Few-layer biotite and van der Waals heterostructures characterization*

Scanning electron microscopy (SEM) images of a multilayer exfoliated biotite flake were collected using a JEOL-JSM 7800F system operated at an accelerating voltage of 10 kV with a lower electron detector (LED). The energy-dispersive spectroscopy (EDS) data acquisition was carried out with an OXFORD detector coupled to the SEM.

The SINS experiments, performed on an ultrathin exfoliated flake of biotite atop Au/Si substrate, were carried out at the Advanced Light Source (ALS) - beamline 2.4 - using a commercial scattering scanning near-field optical microscopy (s-SNOM) instrument (Neaspec GmbH) equipped with a KRS-5 beamsplitter in conjunction with a customized Ge:Cu photoconductor covering the 330-4000cm$^{-1}$ spectral window with 2.5 cm$^{-1}$ spectral resolution [64]. The spectra were collected by integrating over 2048 points with 20.1 ms integration time per point, and they were normalized by a reference spectrum taken at gold surface. The broadband images presented in this work represent the amplitude $|S_2|$ of the s-SNOM signal that carries the local reflectivity response of the material. These images were taken with 200 pixels × 40 pixels with a 28 ms integration time per pixel. All SINS data presented here were taken at the second harmonic of the tip natural resonance frequency for far-field background suppression.

The topography images for evaluating the aging and thermal stability of biotite were acquired by AFM using an Icon Bruker AFM system in standard tapping mode configuration. For the dielectric breakdown experiments, we performed c-AFM measurements using a ParkSystems NX10 microscope, acquiring current-voltage (I-V) curves at different sample positions with a ramp of applied potential difference from 0 to 10 V with 0.1 V/s. A Pt-Ir coated conductive tip from Nanosensors was used to probe a FL-biotite flake on Au/Si substrate grounded by a 1 GΩ resistor using silver ink. Topography and current images with 128 px x 128 px were also acquired in contact mode by applying 30 nN of tip-force and 2 V of potential bias.

The PL and magneto-PL measurements at 3.6K were performed using a helium closed-cycle cryostat with superconducting magnet coils (Attocube - Attodry1000) with magnetic (B) fields up to 9 T applied perpendicular to the vdWH substrate. The sample was mounted on Attocube piezoelectric *xyz* translation stages to control the sample position. The optical measurements were performed using a continuous-wave (CW) laser with a photon energy of 1.88 eV. The PL signal was collimated using an aspheric lens (NA = 0.64) and the selection of circular polarization components was performed before to be focused into a 50 µm multimode optical fiber, being dispersed by a 75 cm spectrometer and detected by a silicon CCD detector (Andor, Shamrock/iDus).

All optical images during BP degradation analysis were acquired using a commercial Nikon optical microscope system. Confocal laser scanning microscopy (CLSM) was performed using a commercial Keyence VK-X200 system with a 150x objective (NA = 0.95) and 408 nm wavelength laser. This technique highlights the BP degradation over time better than standard optical microscope systems due to its higher resolution.

All Raman experiments were done in a WITec Alpha 300R system under a 532 nm excitation laser and a 100x objective (NA = 0.9). To monitor the BP degradation, we used ~500 µW laser power, and the spectrum was acquired with 3 accumulations of 20 sec. For the biotite Raman spectra, the laser was ~10 mW, and the spectrum was acquired with 5 accumulations of 60 sec.

**3 – Results and Discussion**

### *3.1 – Mineralogical identification and characterization*

Fig.1a brings a picture of the geological specimen of biotite used to produce the ultrathin samples investigated in this work. Biotite has a generic chemical formula $K(Mg,Fe)_3(AlSi_3)O_{10}(OH)_2$, forming a solid solution series with the phyllosilicate phlogopite $(KMg_3(AlSi_3)O_{10}(OH)_2)$ [15,46,52]. Both minerals belong to the trioctahedral mica group and differ from each other by the iron (Fe) concentration in their structures: phlogopite is a hydrated phyllosilicate without significant amount of Fe content, while biotite is a Fe-rich crystal [52,65,66]. In this series, the Fe-endmember is an annite mineral with chemical formula $KFe_3AlSi_3O_{10}(OH)_2$. Biotite is formed by two tetrahedral silicon oxide layers (T) with one aluminum substitution intercalated by a magnesium octahedral layer (Oc) in addition to potassium cations between this T-Oc-T stacking oriented along the *c* axis [15,52]. For biotite, Fe atoms are expected to be present in different valence states $(Fe^{2+/3+})$ in the octahedral sites and as $Fe^{3+}$ substitutional ions in the aluminum T sites [67,68]. Due to its lamellar structure with very weak van der Waals forces between layers and strong in-plane bonds [66], biotite has a high-quality basal cleavage and can be easily reduced to FLs by micromechanical exfoliation technique, which we employ in this work. The natural origin of our biotite sample requires a prior identification of mineralogical phases at the nanoscale before understanding its fundamental properties and interaction with other LMs, since heterogeneous elemental compositions and distinct mineralogical phases can coexist in the bulk crystal, replicating in the exfoliated flakes. In this sense, we begin by examining the composition of our natural crystal. From a specific region of a biotite multilayer flake (Fig. 1b), EDS was used to qualitatively evaluate the chemical composition and elemental spatial distribution of our sample, as shown in Figs. 1c-h. The constituent elements aluminum (Al), potassium (K), magnesium (Mg), oxygen (O), and silicon (Si) are observed homogeneously distributed. No impurities were observed other than Fe, also homogeneously distributed and highly incorporated. Although this qualitative elemental analysis guarantees the reproducibility of the composition homogeneity of the exfoliated flakes, it does not make any attribution regarding the mineralogical phase within the biotite-phlogopite series.

A powerful technique to retrieve the mineralogical phase of our sample in its FL form is SINS [62]. Phyllosilicates are well-known by their vibrational assignments and can be identified using standard spectroscopy techniques like Raman and Fourier-transform infrared spectroscopies, as well as by X-ray diffraction analysis [42,69–71]. However, these techniques are most suitable for bulk samples rather than for FL-phyllosilicate flakes due to the low amount of material and/or limited resolution, as discussed in Fig. S1. SINS has then demonstrated to be an alternative tool capable of probing the vibrational modes of ultrathin phyllosilicate minerals with nanometric resolution and sensitivity to layer number [42,62,72–74]. Thus, we performed a vibrational characterization of our ultrathin biotite crystal using SINS. Fig. 1i shows the topography image of a selected staircase-like biotite flake exfoliated atop Au/Si substrate acquired by AFM and its height profile taken at the highlighted white line. The height profile reveals biotite steps varying from 5 up to 15 layers in thickness (considering 1 nm as 1L thickness). Fig. 1j shows the phase signal of the second harmonic demodulated SINS point-spectra $\varphi_2(\omega)$ acquired at different biotite steps as indicated in the infrared (IR) broadband reflectivity near-field image (top panel). The phase spectra are related to the sample absorption and reveal intensity modulation according to flake thickness with absorption bands located approximately at 495, 523, 668, 697, 826, 1049 and 1085 $cm^{-1}$. The observed IR absorption bands for the ultrathin biotite flake are in good agreement with several spectra reported for bulk biotite [75–77]. The band at 495 $cm^{-1}$ is assigned as Si-O-Si bending, while the band at 523 $cm^{-1}$ is assigned as Si-O stretching perpendicular to the tetrahedral plane [76]. The bands at 668 and 697 $cm^{-1}$ are related to Si/Al-O-Si stretching modes. The band at 826 $cm^{-1}$ is related to Al-O stretching vibrations perpendicular to the T plane [75–77]. The higher frequency bands at 1049 and 1085 $cm^{-1}$ are assigned as Si-O-Si stretching modes [75–77]. The contribution of Fe impurities in the spectra is the broadening of peaks due to atomic irregular arrangements caused by the substitutions [75]. Our SINS analysis unambiguously determines biotite as the main mineralogical phase of our FL samples. Moreover, SINS measurements revel that biotite preserves its natural vibrational identity even when the material is thinned down to FLs.

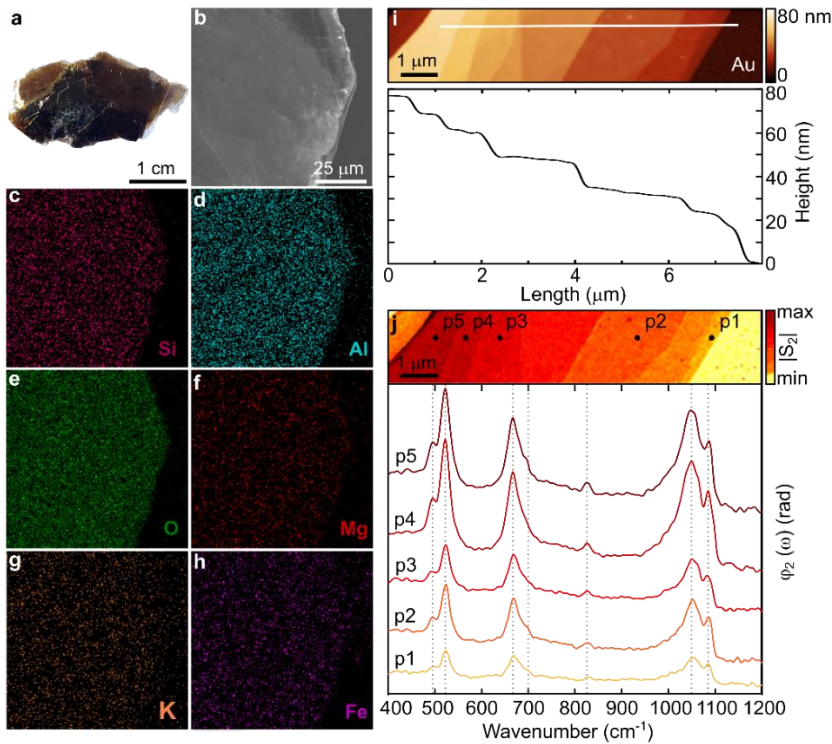

**Figure 1 -** Mineralogical identification and characterization of ultrathin biotite. a) Photography of the biotite bulk-crystal (mine rock) used in this work. b) SEM image of biotite flake edge with the c-h) EDS elemental mappings. i) Height profile extracted along the white line in the AFM topography image (top panel) of a FL-biotite atop Au/Si substrate (bottom panel). j) SINS phase point-spectra $\varphi_2(\omega)$ acquired at different biotite plateaus as indicated in the IR broadband reflectivity near-field image (top panel).

### 3.2 – Stability and electrical characterization of few-layer biotite

After identifying and characterizing the mineralogical attribution of our biotite sample in its FL form, we can shed light into the inspection of its stability and dielectric properties. First, we investigate the aging stability of a FL-biotite sample left in ambient conditions for at least 12 months. Figs. 2a,b present the AFM topography image and height profile of the same flake freshly exfoliated and after 12 months exposed to air, respectively. Our AFM measurements indicate negligible changes on flake surface. For instance, no signs of material swelling, and the root mean square (rms) roughness measured are similar, i.e., $R_{rms}$= 0.26 nm (fresh) and $R_{rms}$= 0.29 nm (after 12 months). These analyses demonstrate that biotite layers are highly stable under open environment.

We also investigate the thermal stability of a FL-biotite flakes. Figs. 2c,d show the topography image of a selected staircase-like biotite flake freshly exfoliated and after performing thermal annealing in air on a hot plate at 300°C for 2 hours, respectively. We highlight regions with thicknesses of 4L to 1L in the topography image of an ultrathin biotite. By comparing both topography images, we observe that the material surface gets cleaner with thermal annealing. However, large clusters of residues (most likely glue from the exfoliation tape) are not entirely removed. Nevertheless, this procedure demonstrates that even a 1L-biotite is thermally stable, supporting thermal treatments without damaging its surface up to 300°C. The respective height profiles before and after thermal annealing for the 1L-biotite region are depicted in Figs. 2c,d as insets. No significant changes in height are observed after the procedure. Moreover, the height profiles show well-defined steps of about 1 nm, confirming the identification of a 1L-biotite with a flat surface and reporting no layer expansion after thermal annealing.

Regarding the electrostatic stability of biotite crystals, the material possesses large electronic bandgap [15,46] and, therefore, can be eventually applied in field-effect transistors with other LMs [16,34,41,78–80]. Thus, an important property in this case is the dielectric breakdown of biotite crystals. For this, we have systematically investigated it by using c-AFM. Here we apply a potential difference through FL-biotite flakes with different thickness, and we measure the maximum potential before the dielectric breakdown ($V_{BD}$). Fig. 2e thus plots I-V curves for biotite capacitor-like structures formed by the metallic coated tip and Au/Si substrate as capacitor plates with a FL-biotite flake with different thicknesses as dielectric media. We do not measure any current signal until it reaches a $V_{BD} \sim 1$ V/nm, which is like reported dielectric breakdown for hBN [81,82], and other phyllosilicates like micas [39,83,84], talc [16,41] crystals, and other LMs [17,29,39]. Figs. 2f,g show the topography and current signal images, respectively, of a biotite flake with thickness of 7 nm acquired simultaneously by c-AFM under 2 V bias from which we can clearly observe that the flake region is well-defined and dark in the current signal image, demonstrating its high insulating capacity below the $V_{BD}$, while the metallic substrate region conducts current. Therefore, these results demonstrate that biotite can indeed be considered a good insulating LM for ultrathin devices applications.

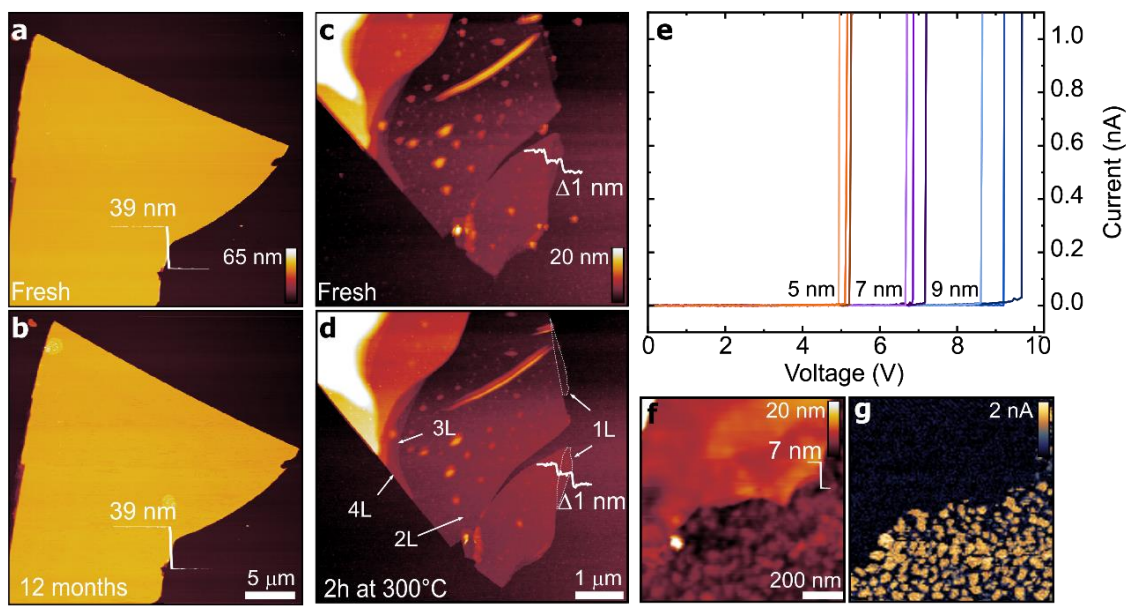

**Figure 2 –** Stability of few-layer dielectric biotite. a) AFM topography image of a selected biotite flake after mechanical exfoliation atop $SiO_2$/Si substrate and b) after 12 months in air with their height profiles at edge as insets. c) AFM topography image of a selected stair-like biotite flake freshly exfoliated atop $SiO_2$/Si substrate and d) after thermal annealing in air on a hot plate at 300°C for 2h. e) I-V curve at three representative positions for biotite flakes with thickness of 5 nm (orange shaded curves), 7 nm (purple shaded curves) and 9 nm (blue shaded curves). f) Topography and g) current signal images of a biotite flake with thickness of 7 nm exfoliated atop Au/Si substrate acquired simultaneously by c-AFM under 2 V bias.

Thus far, we have demonstrated that biotite can be exfoliated down to 1L, presenting an atomically flat surface and acting as a suitable insulating LM. We have also shown its aging and thermal stability while preserving its optical and vibrational properties. In this scenario, two possibilities emerge for the use of biotite in combination with other LMs in vdWHs. Biotite can be integrated in vdWHs either as a layered dielectric substrate or as a capping (encapsulating) layer to protect sensitive LMs, as we present next.

### *3.3 – Few-layer biotite as dielectric substrate layer for optical materials*

Fig. 3a shows the typical optical image of a 1L-MoSe$_2$/biotite/SiO$_2$/Si sample studied in this work, while Fig. 3b brings a typical PL spectrum of the 1L-MoSe$_2$/biotite collected at 3.6 K. The X emission is observed at ~1.661 eV and the trion (T) emission at ~1.629 eV. We observe a T binding energy at ~32 meV which is slightly (~4 meV) higher than the typical values observed for 1L-MoSe$_2$ onto SiO$_2$ or hBN [28,85]. In addition, the X and T emissions have a typical full width at half maximum (FWHM) of ~6 meV which is comparable to the typical values of FWHM observed for 1L-MoSe$_2$/hBN samples [85]. Considering the flake thickness selected in the PL experiments (30-60 nm), our results suggest that such biotite thickness is enough to improve the optical quality of the transition metal dichalcogenides (TMDs) monolayers and could be used in more complex experiments such as gated-PL and others.

We have then investigated the magneto-PL properties of a 1L-MoSe$_2$/biotite sample under a perpendicular B field. Fig. 3c shows the color code map of the circularly resolved PL intensity as a function of the perpendicular B field. The laser excitation is linearly polarized, and the PL detection is σ$^-$ for positive B field. Fig. 3d shows typical circularly polarized PL spectra at B= +9 T and B= -9 T. From this image, we observe that the T peak is strongly circularly polarized at B= 9T which is similar to previous results in the literature [85]. Additionally, Figs. 3e,f show the resulting B field-dependent valley Zeeman splitting (ΔE) fitted (red solid curves) using [28,85,86] $\Delta E = E_{\sigma+} - E_{\sigma-} = g \, \mu_B \, B$, where $\mu_B = 58$ μeV/T is the Bohr magneton, B is the applied perpendicular magnetic field, and g is the valley g-factor, $E_{\sigma+}$ and $E_{\sigma-}$ are the peak energies of the σ$^+$ or σ$^-$ polarized PL of the T or X.

For 1L-TMDs, it is known that the g-factor values depend on the type of excitonic complex, presence of strain field, doping and magnetic proximity effects [28,85,87]. In our data, the obtained values for X and T are $g^X$= -3.1 and $g^T$= -3.4, respectively. We thus remark that these values are lower than typical values reported in literature, usually in the range of -3.8 – -4.2 [85,86,88–91] (see Fig. S2). The reduced values of g-factors could indicate a biaxial compressed strain of order of 4% [89]. However, our Raman and PL results do not evidence any significant strain effect (see Fig. S3). Therefore, a possible explanation could be related to the hybridization of 1L-MoSe$_2$ and biotite interface which may contain magnetic impurities due to its natural occurrence [26]. However, further studies would be necessary to understand in detail the observed magneto-PL results.

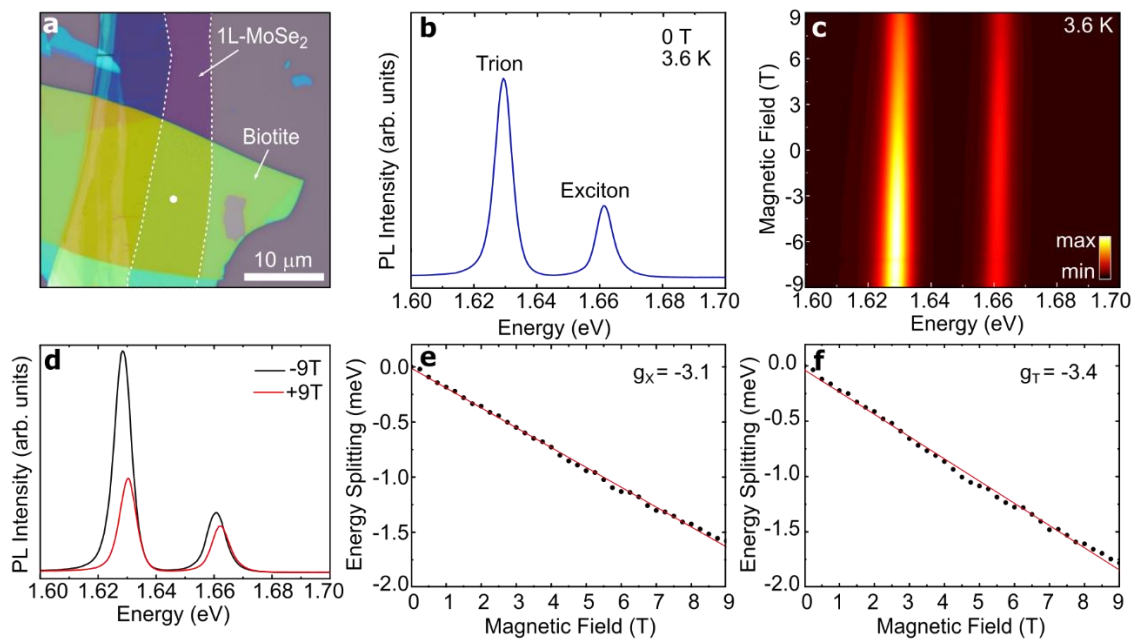

**Figure 3 –** Magneto-PL in 1L-MoSe₂/biotite heterostructures. a) Optical microscope image of 1L-MoSe₂/biotite sample onto SiO₂/Si substrate. Arrows highlight the 1L-MoSe₂ and biotite regions. b) Typical PL spectrum in the 1L-MoSe₂/biotite region at 3.6K collected at the position marked by the white dot in (a). c) Color code map of the circularly resolved PL intensity as a function of the perpendicular magnetic (B) field for linearly polarized optical excitation of 1.88 eV at 3.6K. The PL detection is σ− for the positive B field. d) Circularly polarization resolved PL spectra under applied out-of-plane B field +9 T (black curve) and -9 T (red curve). The exciton (X) PL energy peak is observed at ~1.661 eV while the trion (T) peak at ~1.629 eV. (e) and (f) Exciton and trion energy splitting as a function of out-of-plane B field. The solid red lines are linear fits to the data. The extracted valley g-factor are also indicated in the figures (e) and (f).

### 3.4 – Few-layer biotite as protecting layer for sensitive materials

Now, we demonstrate that biotite can also be used as an encapsulating layer to protect sensitive LMs from environment degradation. Fig. 4a shows an optical microscope image of a BP flake capped by a biotite top layer (dashed white line), as well as BP flakes off from the biotite layer. The BP stability over time (up to 60 days) was evaluated by optical microscopy (Figs. 4a-e) and confocal laser scanning microscopy (CLSM, Fig. 4f). One can see that the BP flakes not covered by the top biotite layer undergo an oxidation process [92–95]. The degradation process can be identified by the formation of bubbles (Fig. 4b-f), changing in color (i.e., BP thickness), followed by a complete deterioration of the material. It is known that BP degradation rate is directly dependent on its thickness [96]. Thus, to ensure that the stability of capped and uncapped flakes here are comparable, the monitoring was systematically carried out with a representative collection of flakes, as depicted in Fig. S3. The BP degradation is further corroborated by Raman spectroscopy, as shown in Figs. 4g,h. While the Raman spectra acquired at the biotite-protected BP locations (black crosses) do not show significant changes over time (Fig. 4g), the uncapped areas (red crosses) barely show Raman peaks associated with BP modes after 60 days (Fig. 4h) [97]. Consequently, this indicates that the biotite layer can undoubtedly protect sensitive LMs from oxidation/degradation processes.

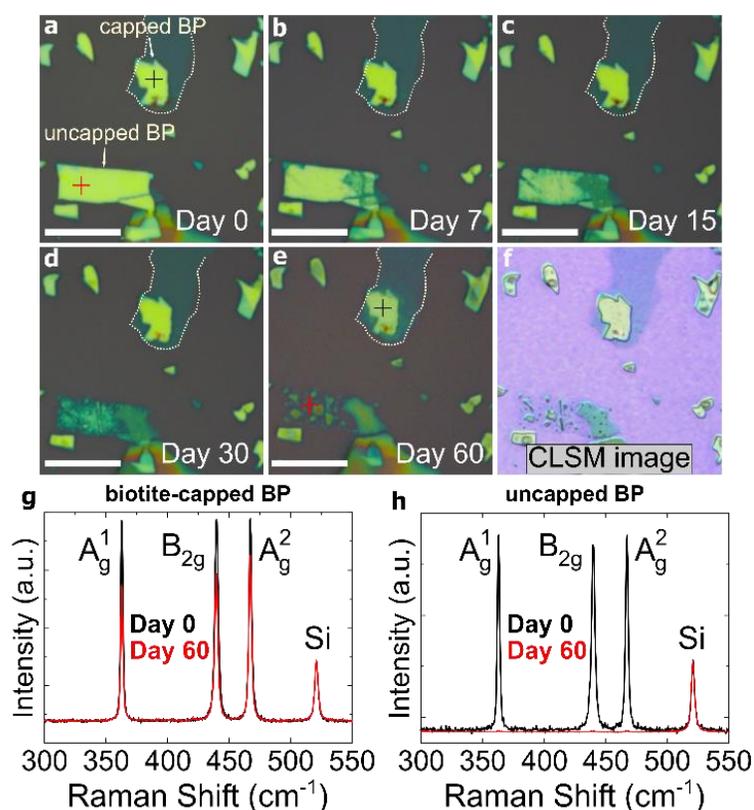

**Fig. 4.** Biotite as protecting layer for black phosphorus (BP). a-e) Optical microscope images of biotite-capped and uncapped BP flakes after the sample was exposed to an open environment for different periods. Scale bar corresponds to 15 μm. f) CLSM image of the same sample region after 60 days of exposure. g) Raman spectra of the biotite-capped and h) uncapped BP flakes onto $SiO_2/Si$ substrate with 0 days (black) and after 60 days (red) under environmental conditions. The spectra are normalized by the Si peak for better comparison. The sample locations where the data were taken are indicated by the red and black crosses in figures (a) and (e).

**4 - Conclusion**

In summary, we presented a comprehensive investigation of ultrathin biotite crystals as a promising naturally occurring dielectric LM for embedding in low-cost LM-based nanodevices. We highlighted its mineralogical origin and elemental distribution through EDS elemental mappings. We used advanced nanoprobe techniques, including synchrotron-based infrared nanospectroscopy, to analyze exfoliated biotite flakes with nanometric resolution and layer number sensitivity to determine their mineralogical phase and vibrational assignments in the IR region. This level of analysis is unattainable using conventional far-field IR absorption techniques due to diffraction limitations at the micrometer scale. Moreover, we demonstrated that FL-biotite obtained by micromechanical exfoliation exhibits high stability to temperature and in open environments. Through biotite dielectric breakdown measurements, we showed that this mineral can be a naturally abundant alternative insulator to synthetic materials like hBN, supporting up to 1 V/nm without conducting current. By combining the environmental, thermal, and electrical stability of biotite, we presented how this mineral can be effectively applied in its FL form as a substrate or capping layer vdWHs. Specifically, we find that ultrathin biotite flakes enhance the quality of the emission in 1L-$MoSe_2$/biotite region and protect sensitive LMs such as BP flakes from degradation for up to 60 days under ambient conditions. We also observe that the g-factors for the 1L-$MoSe_2$/biotite samples are reduced as compared to 1L-$MoSe_2$/$SiO_2$. This effect is not understood yet, but it could be related to the magnetic proximity effect as the biotite has magnetic impurities. Finally, our findings highlight ultrathin biotite as a readily available option of low-cost dielectric material to be applied in ultrathin LM-based applications.

**Declaration of Competing Interest**

The authors declare that they have no known competing financial interests or personal relationships that could have appeared to influence the work reported in this paper.

**Acknowledgments**

All authors thank the financial support from CAPES, CNPq and the Brazilian Nanocarbon Institute of Science and Technology (INCT/Nanocarbono) and thank Professor Marco A. Fonseca from Federal University of Ouro Preto for supplying the biotite crystal. All authors are also thankful to Advanced Light Source (ALS) for SINS measurements, Hans A. Bechtel and Stephanie G. Corder for the experimental assistance. Besides the Brazilian Nanotechnology National Laboratory and Brazilian Synchrotron Light Laboratory, both part of the Brazilian Centre for Research in Energy and Materials (CNPEM), a private non-profit organization under the supervision of the Brazilian Ministry for Science, Technology, and Innovations (MCTI), for sample preparation and characterization – LNNano/CNPEM (Proposals: 20221047, 20221411, 20230017 and 20230147) and LAM (Proposals: 20221265, 20221645, and 20230170) at LNLS/CNPEM. Carlos A. R. Costa is also acknowledged for the AFM experimental assistance. I.D.B., R.O.F., Y.G.G. and A.R.C. acknowledge the support from CNPq (311327/2020-6, 309946/2021-2, 306971/2023-2, and 309920/2021-3). The authors acknowledge Prof. Marcio Daldin Teodoro and Fapesp (grants: 2014/07375-2, 2015/13771-0 e 2022/10340-2) for the use of his laboratory for magneto-optical measurements. C.J.S.de.M. acknowledges the financial support of FAPESP (2022/11526-2) and MackPesquisa. I.D.B. and A.R.C. acknowledge the FAPESP financial support (2022/02901-4), Y.G.G. thanks the FAPESP grant (22/08329-0), as well as I.D.B. and R.O.F (Young Investigator Grant 2019/14017-9). Finally, all authors thank Dr Jessica Fonsaca and Professor Angelo Malachias for enlighten discussions.

**Supplementary Material.**

Supplementary data to this article can be found online.